# The search for empirical formulae for the aftershocks descriptions of a strong earthquake

A.V. Guglielmi

*Schmidt Institute of Physics of the Earth, Russian Academy of Sciences, ul. B. Gruzinskaya 10, Moscow, 123995 Russia, e-mail: guglielmi@mail.ru*

**Abstract**

The paper is based on the report read by the author on October 24, 2018 at the meeting of the Scientific Council of the Institute of Earth Physics of the Russian Academy of Sciences. The report was dedicated to the 150th anniversary of the outstanding Japanese seismologist Fusakichi Omori. As is known, Omori established the first empirical law of the earthquakes physics, bearing his name. The Omori law states that the frequency of aftershocks on average decreases hyperbolically over the time. Three versions of Omori law are described briefly. The recent version allows to poses the inverse problem of the earthquake source, that "cools down" after the main shock.

**Table of contents**

## 1. Introduction

On October 24, 2018 at the meeting of the Academic Council of the Institute of Physics of the Earth RAS the author made a report dedicated to the 150th anniversary of the outstanding Japanese seismologist Fusakichi Omori (1868 – 1923). The presented paper summarizes the contents of this report.

While still quite young, at the age of 26, Omori made an outstanding contribution to science, which has not lost its value these days [Davison, 1924; Guglielmi, 2017]. He discovered the hyperbolic dependence of the frequency of aftershocks on time. And this was the first empirical law of the physics of earthquakes.

It is clear that observations are the main source of our knowledge, and we also understand that the selection of empirical formulas plays a key role in systematizing our experience. Sometimes the empirical formula turns out to be fundamental. For example, Ohm's law lies in the fundamentals of electrical engineering and geoelectrics. Another example is the formula for displacement current. Recall that the formula was selected by Maxwell under the strong pressure of experienced facts. It is not superfluous to mention Planck's famous formula describing the emission spectrum of the absolutely black body. This formula was also chosen empirically. Finally, let us recall the Hubble's law of expansion of the Universe, discovered by observing the motion of galaxies.

Let us turn, however, to our history. It began in 1894, when Omori published the result of his research [Omori, 1894]. This event was preceded by an interesting prehistory. In 1850 John Milne destinated to become one of the founders of the modern seismology was born in Liverpool [Davison, 1927]. In 1880, Milne created a horizontal pendulum seismograph, which proved to be easy to use and sensitive enough device for recording earthquakes. At the end of the century before last, Milne seismographs recorded strong earthquakes in Japan. Fusakichi Omori analyzed the seismograms and discovered the hyperbolic law of lowering the frequency of aftershocks over the time.

In this paper we describe the Omori law [Omori, 1894], consider its generalizations [Hirano, 1924; Utsu, 1961; Utsu, Ogata, Matsu'ura, 1995; Guglielmi, 2016a,b] and formulate the inverse problem of the earthquake source [Guglielmi, Zavyalov, 2018]. Solving the inverse problem gives us the opportunity to approach aftershock processing and analysis in a new way [Guglielmi, Zotov, Zavyalov, 2018].

## 2. Three wordings of the Omori law

In its original formulation, the Omori law is as follows:

$$n(t) = \frac{k}{c+t}. \tag{1}$$

Here $n$ – aftershock frequency, $k > 0$, $c > 0$, $t \geq 0$ [Omori, 1894]. The introduction of the parameter $c$ is partly due to the technical difficulties of registering repeated shocks immediately after the main shock of the earthquake. In essence, the Omori formula is one-parameter. The parameter $k$ gives information about the rate of decay of aftershock activity. It is the most important integral characteristic of the earthquake source that "cools" after the main shock.

For a long time after the discovery, formula (1) was not given special attention. The interests of seismologists focused on the propagation of seismic waves, on the study of internal structure of the Earth by seismology methods, and on the problem of the origin of earthquakes. Particular attention was paid to the development of earthquake resistant design principles, the development of methodological foundations of mineral exploration, and the search for precursors of catastrophic earthquakes. Considerable efforts were also directed at improving recording devices, at creating a network of seismological stations, and at drawing up detailed maps and catalogs of earthquakes. These efforts made it possible to achieve significant success in studying the spatial-temporal distribution of seismic events [Davison, 1927].

Only in 1924, i.e. the next year after Omor's death, Hirano selected another empirical formula for approximating the frequency of aftershocks [Hirano, 1924]:

$$n(t) = \frac{k}{(c+t)^p} \qquad (2)$$

The Hirano formula, unlike the Omori formula, is two-parameter, which in itself, of course, improves the approximation of the experimental data.

It is difficult to say why formula (2) did not immediately enter into the practice of processing and analyzing aftershocks. Perhaps in part because Hirano proposed to select different values of $p$ for different time intervals within the same series of aftershocks. However, the formula $n = k/(c+t)^{p(t)}$ is not quite correct from the point of view of a mathematical orthography. (Perhaps that is why Jeffreys preferred to use formula (1), rather than (2) in his study of aftershocks [Jeffreys, 1938].) Utsu proposed to consider $p = \text{const}$ for each particular aftershock series [Utsu, 1961], and since then, the power function (2) is widely used in seismology. The parameter $p$ varies from case to case approximately in the range from 0.7 to 1.5 (see reviews [Utsu et al., 1995; Guglielmi, 2017] and the literature indicated in them).

The third version of the Omori law was found on the basis of analogy between the recombination of charged particles in the ionosphere after sunset and the deactivation of the earthquake source, "cooling down" after the main shock [Guglielmi, 2016a]. It has the form of the differential equation

$$\frac{dn}{dt} + \sigma n^2 = 0. \qquad (3)$$

It is easy to verify that if $\sigma = 1/k$ the general solution of equation (3) coincides with the empirical Omori formula. The parameter $\sigma$ will be called the deactivation coefficient of earthquake source. The value of $\sigma$ indicates how fast the earthquake source loses its ability to excite aftershocks [Guglielmi, 2016b].

The advantage of writing the Omori law in the form of differential equation is quite obvious. Indeed, the equation (3) tells us nontrivial generalizations of the law of aftershocks evolution. For example, we can take into account the space-time distribution of aftershocks by adding a diffusion term to equation (3), i.e. write it as

$$\partial n / \partial t = -\sigma n^2 + \hat{D}\nabla^2 n, \qquad (4)$$

where

$$\hat{D} = \begin{Vmatrix} D_\parallel & 0 \\ 0 & D_\perp \end{Vmatrix}. \tag{5}$$

Here we have taken into account the anisotropy of faults in the earth crust at the phenomenological level [Guglielmi, 2017; Zotov, Zavyalov, Klain, 2018].

Another important generalization is motivated by the following considerations. On the one hand, we do not want to introduce additional phenomenological parameters (for example, *p*) that do not have a clear geodynamic sense. Those, we want to leave the law of evolution one-parameter. On the other hand, a rather complex relaxation process begins in the earthquake source after the main shock, the geological environment becomes non-stationary, and we would like to take this non-stationarity into account. It is quite clear that one cannot simply put the parameter *k* time-dependent in the Omori formula (1) for the same reason that the parameter *p* in the Hirano-Utsu formula (2) cannot be considered time-dependent. But nothing prevents us from considering the time-dependent deactivation coefficient in equation (3): $\sigma = \sigma(t)$. Then the general solution of evolution equation takes the form

$$n(t) = n_0 \left[ 1 + n_0 \int_0^t \sigma(t') dt' \right]^{-1}. \tag{6}$$

If $\sigma = \text{const}$, then, up to redefinition formula (6) coincides with the Omori formula (1). If $\sigma \neq \text{const}$, then the formula (6) flexibly simulates time-dependent relaxation of the earthquake source to a new metastable state.

In conclusion of this section we briefly discuss one of the ways for the further development of the theory. Let us use the distant analogy between the variations of the state of rocks in the earthquake source and the variations of the Earth's climate. Suppose that there is an equilibrium state $\bar{\sigma}$, generally speaking, depending on time. Let $\tau$ is the characteristic time approached $\sigma$ to equilibrium. Then the relaxation theory of deactivation can be based on the equation

$$\frac{d\sigma}{dt} = \frac{\bar{\sigma}(t) - \sigma}{\tau} + \xi(t), \tag{7}$$

similar to that used in climatology to describe the average temperature on the earth's surface. Function $\xi(t)$ simulates the effect on the earthquake source of endogenous and exogenous triggers.

### 3. Inverse problem

Rewrite the equation (3) as follows

$$\int_0^t \sigma(t') dt' = g(t). \tag{8}$$

Here we have introduced the designation $g(t) = [n_0 n(t)]^{-1}[n_0 - n(t)]$. The auxiliary function $g(t)$ is known experimentally, and equation (8) resembles the Volterra integral equation of the first kind. Of course, the kernel of our equation is trivial. Nevertheless, the structure (8) tells us that we can set and solve the inverse problem, i.e. to find the unknown function $\sigma(t)$, using the known aftershock frequency $n(t)$ [Guglielmi, 2017].

Like many inverse problems, our inverse problem is posed incorrectly. Regularization is smoothing auxiliary function $g(t)$ [Guglielmi, Zavyalov, 2018; Guglielmi, Zotov, Zavyalov, 2018]. After regularization, the solution is $\sigma = d\langle g \rangle / dt$. Here, the angle brackets denote smoothing.

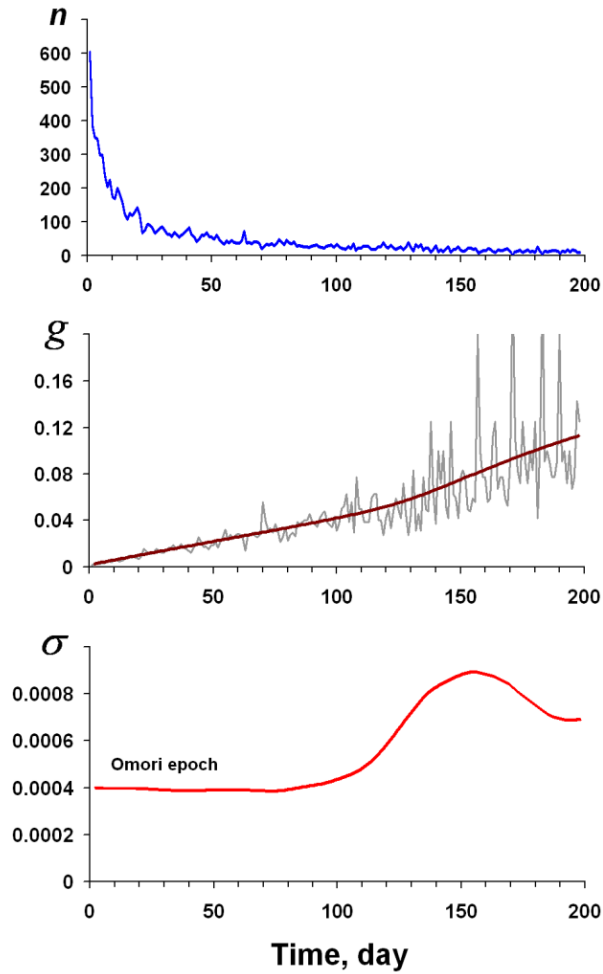

An example of solving the inverse problem. The event took place in Southern California 17.01.1994. The magnitude of main shock is M = 6.7.

The result of aftershock analysis using the method oa inverse problem is shown in the figure. On the top panel we see the frequency of aftershocks. The middle panels show auxiliary functions before and after regularization. The bottom panel shows the function of deactivation. We see that during the first hundred days sigma is constant. It is natural to call such a period of time the epoch of Omori.

The duration of the Omori epoch varies considerably from case to case, and in some cases the Oiori epoch is completely absent. The study found a weak tendency to increase the duration of the Omori epoch with increasing magnitude of the main shock.

The authors of the paper [Guglielmi, Zotov, Zavyalov, 2018] proposed to create an Atlas of aftershocks based on solving the inverse problem of the earthquake source. The authors view the creation of the Atlas as a collective project, and call on interested seismologists to join the project. The processing of aftershocks by the proposed method is no more difficult than by the method of Omori or Utsu. The analysis of information accumulated in the Atlas will give interesting results.

### 4. Conclusion

In honor of the 150th anniversary of Fusakichi Omori, we analyzed various aspects of his discovery and came to a conclusion that the representation of the Omori law in the form of aftershock evolution equation gives us an interesting opportunity to set and solve the inverse problem of earthquake source that "cools" after the main shock.

As a reward for a serious exposition of purely scientific aspects of our problem, I beg you to take into account my own subjective viewpoint on the Omori law. I adhere to hypotheses on the existence of hyperbolicity principle, which was adopted by Omori in formulating the law. From time to time we observe the Omori epoch of varying duration. And does this suggest to us that the fundamental principle of hyperbolicity is fulfilled in aftershock physics, just as in general physics, for example, the principle of positional invariance is strictly fulfilled? Deviations from hyperbolicity can be associated with the effect on the earthquake source of endogenous and exogenous triggers, described in the review [Guglielmi, 2015].

*Acknowledgments*. I express my deep gratitude to A.D. Zavyalov and O.D. Zotov for numerous discussions of the aftershock evolution. Sincerely thank B.I. Klain and A.S. Potapov for interest in this work and for valuable comments. The work was supported by the project of the RFBR 18-05-00096, Program No. 12 of the RAS Presidium, as well as the state assignment program of the IPhE RAS.

**Author Information**

**GUGLIELMI Anatol Vladimirovich** − Prof., Dr. Sci. (Phys.-Math.), Chief Researcher, Institute of Physics of the Earth, RAS, 123995, 10 B. Gruzinskaya 123995, Moscow, Russia. Ph.: +7 (495) 582-99-71. E-mail: guglielmi@mail.ru